\documentclass[sn-mathphys]{sn-jnl}

\jyear{2022}%

\theoremstyle{thmstyleone}%
\theoremstyle{thmstyletwo}%

\theoremstyle{thmstylethree}%

\usepackage{url}
\usepackage{breakurl}
\usepackage{lineno}
\usepackage{amsmath,xspace,upgreek}

\begin{document}

\title[Article Title]{High energy bremsstrahlung at the FCC-ee, FCC-eh and LHeC}


\author*[1]{\fnm{Krzysztof} \sur{Piotrzkowski}}\email{krzysztof.piotrzkowski@agh.edu.pl}

\author*[1]{\fnm{Mariusz} \sur{Przybycien}}\email{mariusz.przybycien@agh.edu.pl}

\affil[1]{\orgname{AGH University of Science and Technology},  \orgdiv{Faculty of Physics and Applied Computer Science},
\orgaddress{\street{Al.~Mickiewicza~30,}~\postcode{30-055}~\city{Krakow}, 
\country{Poland}}}


\abstract{Bremsstrahlung spectra will be  strongly distorted due to small lateral beam sizes at  future colliders. That in turn will have large consequences for the electron and positron beam lifetimes as well as for the luminosity measurements in the case of electron-hadron colliders. We discuss in detail such consequences for the Future Circular Collider and Large Hadron electron Collider cases.}




\maketitle

\section{Introduction}
\label{sec:intro}
In collisions of high energy charged particles, the bremsstrahlung process ("braking radiation") takes place over macroscopic lateral distances, or at extremely large impact parameters. As a result, bremsstrahlung becomes suppressed in high energy collisions of beams with small lateral sizes, as was observed at VEPP-4 \cite{Blinov:1982vp}, HERA \cite{Piotrzkowski:1995xh}, LEP \cite{Burkhardt:1994wk} and KEKB \cite{Kotkin:2005rm}. A very good understanding of bremsstrahlung at colliders is important as on the one hand it is usually the limiting factor for the electron and positron beam lifetimes, and on the other hand, the high energy bremsstrahlung is used for the precise luminosity measurements, of the electron-hadron collisions in particular. Below, the relevance of such {\it Beam-Size Effects} (BSE) is discussed in-depth in the context of the future colliders -- the electron-positron Future Circular Collider (FCC-ee) \cite{FCC:2018evy}, the Large Hadron electron Collider (LHeC) \cite{LHeC:2020van} and the electron-hadron Future Circular Collider (FCC-eh) \cite{FCC:2018vvp}. 

\begin{figure}[tbph]
\centering
\includegraphics[width=0.35\textwidth]{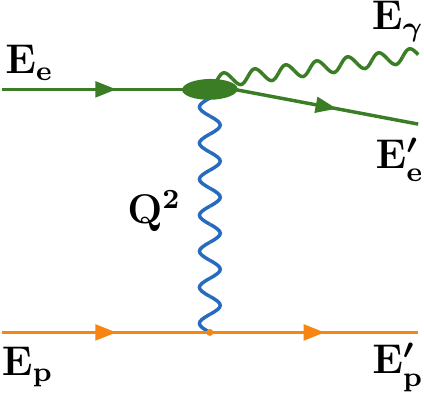}
\caption{Bremsstrahlung diagram for electron-positron or electron-proton collisions. At high energies, the proton structure can be neglected thanks to the long-range nature of this process.}
\label{fig:diag} 
\end{figure}

Macroscopically large impact parameters in bremsstrahlung stem from the extremely small virtuality $Q^2$ of the exchanged photons in this process at high energies (see Fig. \ref{fig:diag}), as its minimal value $Q^2_\mathrm{min}= m_e^2 y^2/ \left( 4\gamma_e\gamma_p(1-y)\right)^2$,
where $m_e$ is the electron mass, $y=E_\gamma/E_e$, $E_\gamma$ is the energy of the radiated photon and $E_e$ the electron beam energy, $\gamma_e=E_e/m_e$ and $\gamma_p$ is the proton Lorentz factor for the electron-proton collisions or the positron one for the electron-positron case. The minimal virtuality $Q^2_\mathrm{min}$ is due to a longitudinal momentum transfer only, without any lateral contribution -- this results in a typical size of lateral momentum transfers being not far from $\vert Q\vert_\mathrm{min}$, as the bremsstrahlung amplitude strongly decreases with increasing $Q^2$. For example, in case of the electron-positron collisions at $\sqrt{s}=365$~GeV, the maximal impact parameter $b_\mathrm{max}$ exceeds 1~m for $y\approx 0.2$, assuming $b_\mathrm{max}={\mathrm \hbar}/\vert Q\vert_\mathrm{min}$. 

The Bethe-Heitler cross section\footnote{It is accurate to well below 1\% at high-energies and for $E_e-E_\gamma \gg m_e$, if the BSE are negligible.} for the electron bremsstrahlung is:
\begin{equation}
    \frac{{\rm d}\sigma_\mathrm{BH}}{{\rm d}y} = \frac{16\alpha^3}{3m_e^2}\frac{1}{y}\left(1-y+\frac{3}{4}y^2\right)L,
    \label{eq:1}
\end{equation}
where $\alpha$ is the fine structure constant, and
$L=\ln{(m_e/\vert Q\vert_\mathrm{min})}-0.5$. For all practical purposes, one can assume that the energy of the scattered electron is equal to $E_e-E_\gamma$, as in the laboratory frame this equality holds with a precision of approximately $\gamma_p\vert Q\vert_\mathrm{min}/E_e$.

At asymptotically large energies, with good accuracy, one can represent the BSE in head-on beam collisions by replacing $\vert Q\vert_\mathrm{min}$ in Eq. \eqref{eq:1} with a constant derived from the lateral beam sizes.  This results in an asymptotic independence of the bremsstrahlung cross section from the beam energies which is closely related to the asymptotic "full screening" effect in the electron-gas bremsstrahlung, where a constant $b_\mathrm{max}$ cut-off is effectively set by the size of atom or molecule.

One should note here, that the BSE are present only if both beams have small transverse sizes, and that it has nothing to do with particle densities -- in fact, the BSE are fully present for collisions of single-particle-beams\footnote{However, for collisions of single particles instead of formulae valid for the Gaussian-shaped particle bunches, used below, one should start from the most general ones -- see Eqs. 4.8--4.10 in Ref.~\cite{Kotkin:1992bj}.}, and are independent of the number of particles in colliding bunches, hence might be called "geometrical" effects. However, the following results are limited to $y>0.001$ as for such bremsstrahlung events their formation or {\it coherence} length (see Ref. \cite{Klein:1998du}, for example), $l_c = 4\hbar c\gamma_e^2/E_\gamma = 4\hbar c\gamma_e/(ym_e)$, is much shorter than the length of interaction area, and below $y=10^{-3}$ the coherent effects might become significant and should be then taken into account.

The results reported below were obtained using the BSE framework described in Ref.~\cite{Kotkin:1992bj}, with the additional effects (as a bremsstrahlung amplification) due to the beam lateral displacements reported in Ref.~\cite{Piotrzkowski:2020uya}.
\section{Beam-size effects at the FCC-ee}
Electron-positron collisions at the Future Circular Collider will be made at four centre-of-mass energies -- 91.2 GeV (at the $Z$-pole), 160 GeV (at the $WW$ production threshold), 240~GeV (for the $WH$ production) and 365 GeV (at the $t\Bar{t}$ threshold). The colliding beams will be very flat with an aspect ratio larger than $200\! :\! 1$, see Tab.~\ref{tab:i} \cite{FCC:2018evy}. The expected observed bremsstrahlung cross-sections are calculated by subtracting from the nominal Bethe-Heitler cross-section a correction due to the BSE: 
$\sigma_\mathrm{obs}=\sigma_\mathrm{BH}-\sigma_\mathrm{corr}$.
\begin{table}[tbph]
\centering
{\renewcommand{\arraystretch}{1.2}
\begin{tabular}{|c|c|c|c|c|}
\hline
$E_\mathrm{beam}$[GeV]& 45.6 & 80 & 120 & 182.5\\
\hline 
$\sigma_x$ [$\upmu$m] & 8 & 21 & 14 & 39\\
$\sigma_y$ [nm] & 34 & 66 & 36 & 69\\
\hline
\end{tabular}}
\ \\[10pt]
\caption{FCC-ee lateral beam-sizes at the interaction points.}
\label{tab:i}
\end{table}

Bremsstrahlung in high energy electron-positron collisions, also called the radiative Bhabha scattering, is identical for electrons and positrons. 

\subsection{Bremsstrahlung spectra}
\begin{figure}[tbph]
\centering
\includegraphics[width=0.75\textwidth]{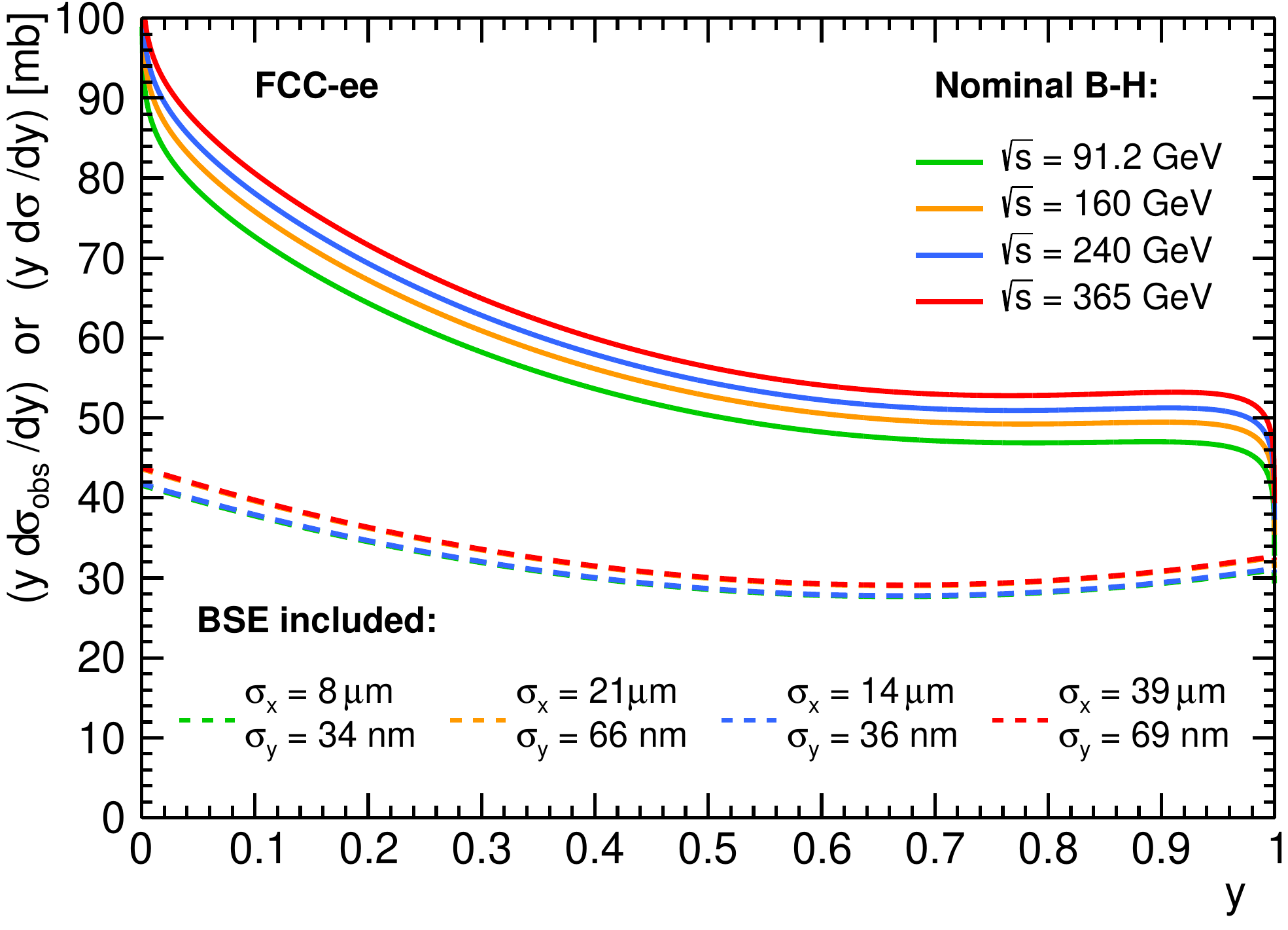}
\caption{Bremsstrahlung spectra at the $\sqrt{s}=91.2$ GeV, 160 GeV, 240 GeV and 365 GeV FCC-ee -- solid lines $y\,{\rm d}\sigma/{\rm d}y$ are for the Bethe-Heitler nominal case and dashed ones $y\,{\rm d}\sigma_\mathrm{obs}/{\rm d}y$ when the BSE is included. Note that the spectra with the BSE included overlap due to similar $\sigma_y$: those at $\sqrt{s}=160$ GeV and 365 GeV, as well as those at $\sqrt{s}=91.2$ GeV and 240 GeV.}
\label{fig:1} 
\end{figure}

\begin{figure}[tbph]
\centering
\includegraphics[width=0.75\textwidth]{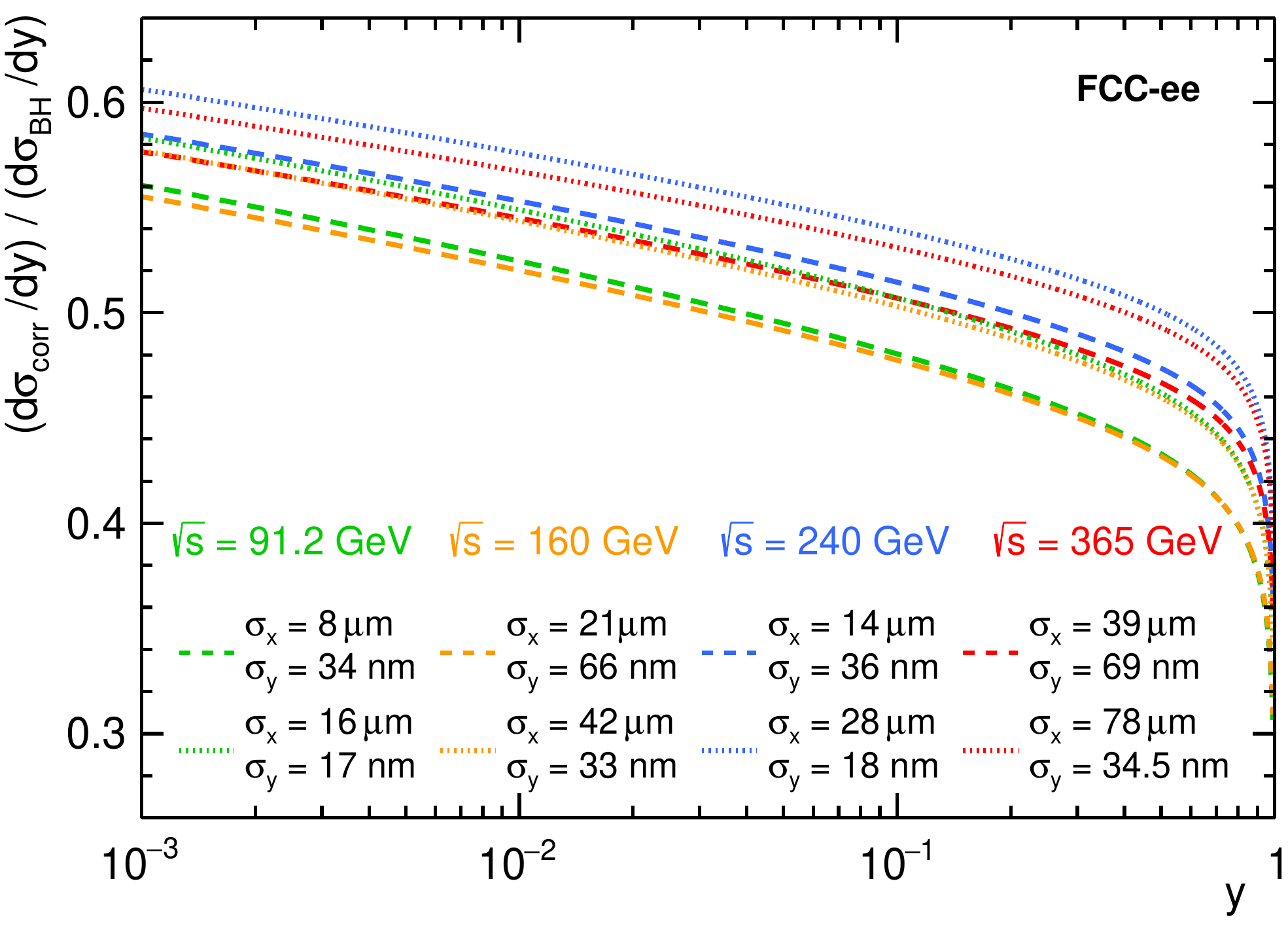}
\caption{Relative size of bremsstrahlung suppression at the $\sqrt{s}=91.2$, 160, 240 and 365 GeV FCC-ee. Dashed lines represent the BSE for the nominal beam sizes and the dotted ones correspond to the BSE case with $\sigma_x \times 2$, $\sigma_y/2$.}
\label{fig:2}
\end{figure}

The bremsstrahlung spectra at the FCC-ee will be very strongly distorted by the BSE, even at the highest photon energies, for $y$ close to 1, as shown in Fig.~\ref{fig:1}. 
The observed bremsstrahlung cross-sections
are very weakly dependent on beam energy and vary significantly only with a change of vertical beam sizes, as expected for the BSE asymptotic regime\footnote{In Appendix \ref{app_1} a simple function is described which provides precise predictions, for a given $\sigma_y$, of the BSE cross-sections at the FCC-ee.}.
At low photon energies, the bremsstrahlung suppression exceeds 50\% -- see Fig.~\ref{fig:2}, and a four-fold increase of the beam aspect ratios, while keeping the transverse beam area constant, would increase the BSE suppression only by about 3\%.

\begin{figure}[tbph]
\centering
\includegraphics[width=0.8\textwidth]{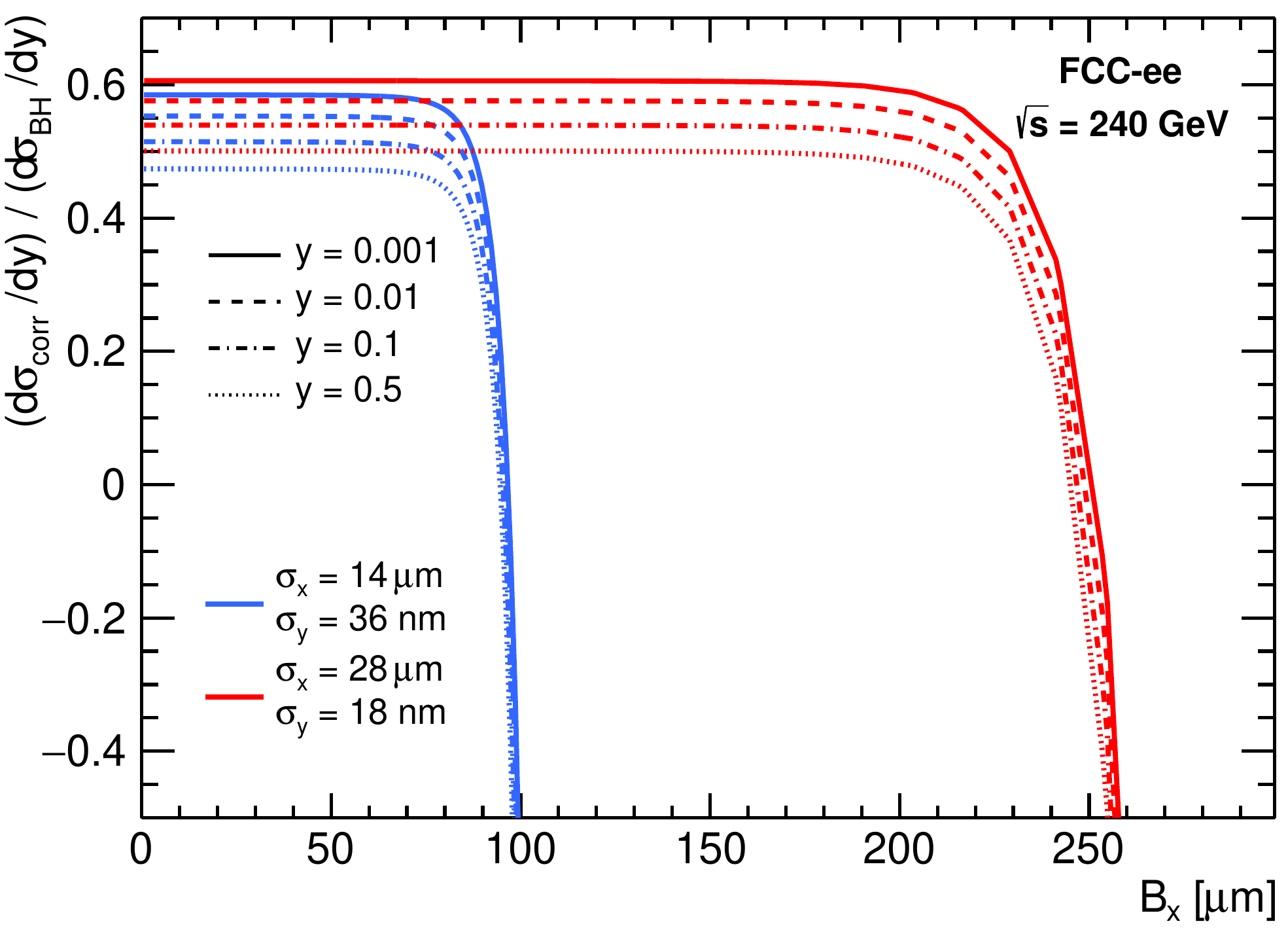}
\includegraphics[width=0.8\textwidth]{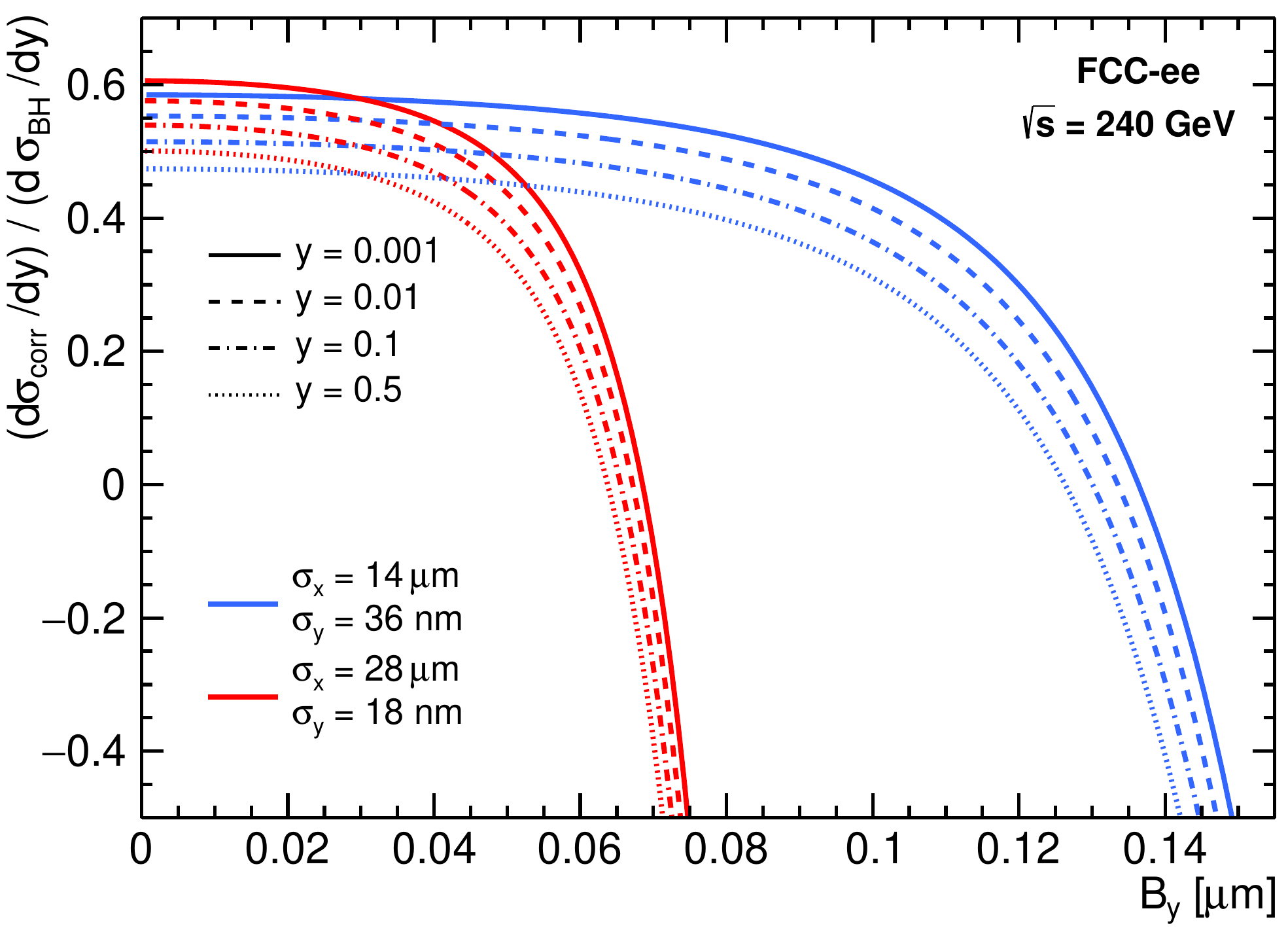}
\caption{Variations of the bremsstrahlung cross-sections at the $\sqrt{s}=240$ GeV FCC-ee for the horizontal (upper plots) and vertical (bottom plots) beam displacements or offsets. The blue lines correspond to the nominal beam sizes and the red ones to the case with a 4 times bigger beam aspect ratio. Results at different values of $y=E_\gamma/E_e$ are shown with different line styles.}
\label{fig:3} 
\end{figure}

The size of BSE varies also with the relative lateral beam displacements -- at large displacements, it is even leading to the effective bremsstrahlung amplification, manifested by negative $\sigma_\mathrm{corr}$. In Fig.~\ref{fig:3} the relative BSE corrections due to the horizontal and vertical beam displacements are shown for the 240 GeV FCC-ee case. These effects are very similar at all four centre-of-mass energies -- significant horizontal displacements do not result in a significant change of the bremsstrahlung cross-section, in contrast to vertical displacements leading to a visible decrease of the suppression even for relatively small (wrt. the corresponding beam size) beam lateral shifts. This overall behaviour does not change much with a significant variation of the beam aspect ratio, and the weak sensitivity of the BSE at the FCC-ee to the horizontal beam offsets implies a small sensitivity also to the horizontal beam crossing angles.

At high energy $e^+e^-$ colliders the non-radiative Bhabha scattering is used for precise absolute luminosity measurements. Since the scattering angles of the detected electrons and positrons are not negligible therefore the exchanged photon virtuality is significant. As a result, the event rates are not affected by the BSE in this case. 

\subsection{Lifetime limits of electron and positron beams}
The beam lifetime $\tau$ is one of the crucial parameters of high energy colliders -- the bigger is $\tau$ the higher is the average collider luminosity for a given initial luminosity. In addition, shorter beam lifetimes require more frequent beam refilling, which might be particularly challenging in the case of the positron beams. 

There are many sources of beam particle losses which are affecting the beam lifetime, but its ultimate, irreducible limit is due to particle scattering at the collider interaction points. Such a loss may occur due to either a significant energy reduction of the beam particle, its annihilation or due to a large scattering angle, as a result of a single $e^+e^-$ interaction. At the FCC-ee the largely dominant particle loss mechanism proceeds via the energy decrease due to the emission of high-energy photons at very small angles, in the bremsstrahlung process.

These particle losses are well modelled by introducing the energy aperture $a$ which is simply the maximal fractional energy decrease of beam particles which is still accommodated by the collider beam optics\footnote{Or, $a$ is the minimal fractional energy decrease for the lost particles.}. At the FCC-ee this aperture varies, from below 0.01 to almost 0.02 for low and high-energy beams, respectively.

The beam lifetime $\tau_b$ due to bremsstrahlung can be then calculated as
$\tau_b = N_b/(L_\mathrm{tot}\sigma_{a})$, where $N_b$ is the nominal number of particles in the "radiating beam", $L_\mathrm{tot}$ is the nominal FCC-ee luminosity (summed over four interaction points) and $\sigma_{a}=\int^1_{a} d\sigma_\mathrm{obs}$, where $a=y_\mathrm{min}$.
\begin{table}[tbph]
\centering
{\renewcommand{\arraystretch}{1.2}
\begin{tabular}{|l||c|c|c|c|}
\hline
$E_\mathrm{beam}$ [GeV]& 45.6 & 80 & 120 & 182.5\\
\hline 
$N_b$ [$10^{13}$] & 242 & 25.5 & 5.05 & 0.945\\
$L_\mathrm{tot}$ [$10^{35}$ cm$^{-2}$s$^{-1}$] & 74.8 & 7.64 & 2.92 & 0.5\\
$\sigma_{a}$ [mb] for $a=0.01 (0.02)$ &  166 (137) & 174 (144)  & 167 (138) & 175 (145) \\ 
$\sigma_\mathrm{BH}$ [mb] for $a=0.01 (0.02)$ &  319 (260) & 333 (271)  & 343 (280) & 353 (288) \\ 
\hline
$\tau_{b}$ [min] for $a=0.01 (0.02)$  & 32 (39) & 32 (39) &  17 (21) & 18 (22)\\
\hline
\end{tabular}}
\ \\[10pt]
\caption{FCC-ee beam lifetimes due to bremsstrahlung, assuming the most recent beam parameters as given in Tab. 1 in Ref. \cite{Zimmermann:2023uqi}.}
\label{tab:ii}
\end{table}

The BSE are very beneficial at the FCC-ee as the ultimate beam lifetimes are about two times bigger thanks to the effective bremsstrahlung suppression due to small lateral beam sizes, see Tab. \ref{tab:ii}. One should note that these are instantaneous beam lifetimes, assuming perfectly Gaussian particle bunches of given sizes, and colliding head-on exactly. The suppression is weakly dependent on the energy aperture as well as on the beam offsets (see Fig.~\ref{fig:3}). Beams will not collide head-on at the FCC-ee -- a full horizontal beam crossing angle of 30~mrad is foreseen \cite{FCC:2018evy}, but that should not affect the bremsstrahlung suppression since its sensitivity to significant horizontal offsets is negligible, as demonstrated by upper plots in Fig.~\ref{fig:3}. 

The FCC-ee beams will be very strongly focused in the vertical plane leading to the {\it hour-glass effect} -- a significant change of vertical beam size during the bunch crossing. Using $R_{HG}$, the luminosity correction factors for the hour-glass effect (see Tab. 2.1 in Ref. \cite{FCC:2018evy}), the effective vertical beam-size $\sigma_{y,e\!f\!\!f}=\sigma_y/R_{HG}$ is evaluated.
The significance of the hour-glass effect to the BSE at the FCC-ee can be then estimated by replacing $\sigma_y$ with $\sigma_{y,e\!f\!\!f}$ what leads to an increase of the observed cross-sections reported above in Tab. \ref{tab:ii} by 0.35\% at the lowest beam energy and up to 1.3\% at the highest one. The corresponding beam lifetimes will then decrease accordingly.
\section{Beam-size effects at the LHeC and FCC-eh}

At electron-hadron colliders bremsstrahlung plays a very important role also because of its use for precise measurements of absolute luminosity \cite{ZEUSLuminosityGroup:2001eva,Piotrzkowski:2021cwj}. However, in this case, the beam size effects become a problem as they need to be accurately corrected for. At HERA~I, the BSE were small and the theoretical predictions could be verified only with a precision of about 30\%~\cite{Piotrzkowski:1995xh}, but the BSE calculations can be profoundly verified at the planned Electron-Ion Collider at Brookhaven, including the surprising effect of bremsstrahlung amplification for large beam offsets \cite{Piotrzkowski:2020uya}.
\subsection{Bremsstrahlung spectra and luminosity measurement}
\begin{figure}[tbph]
\centering
\includegraphics[width=0.75\textwidth]{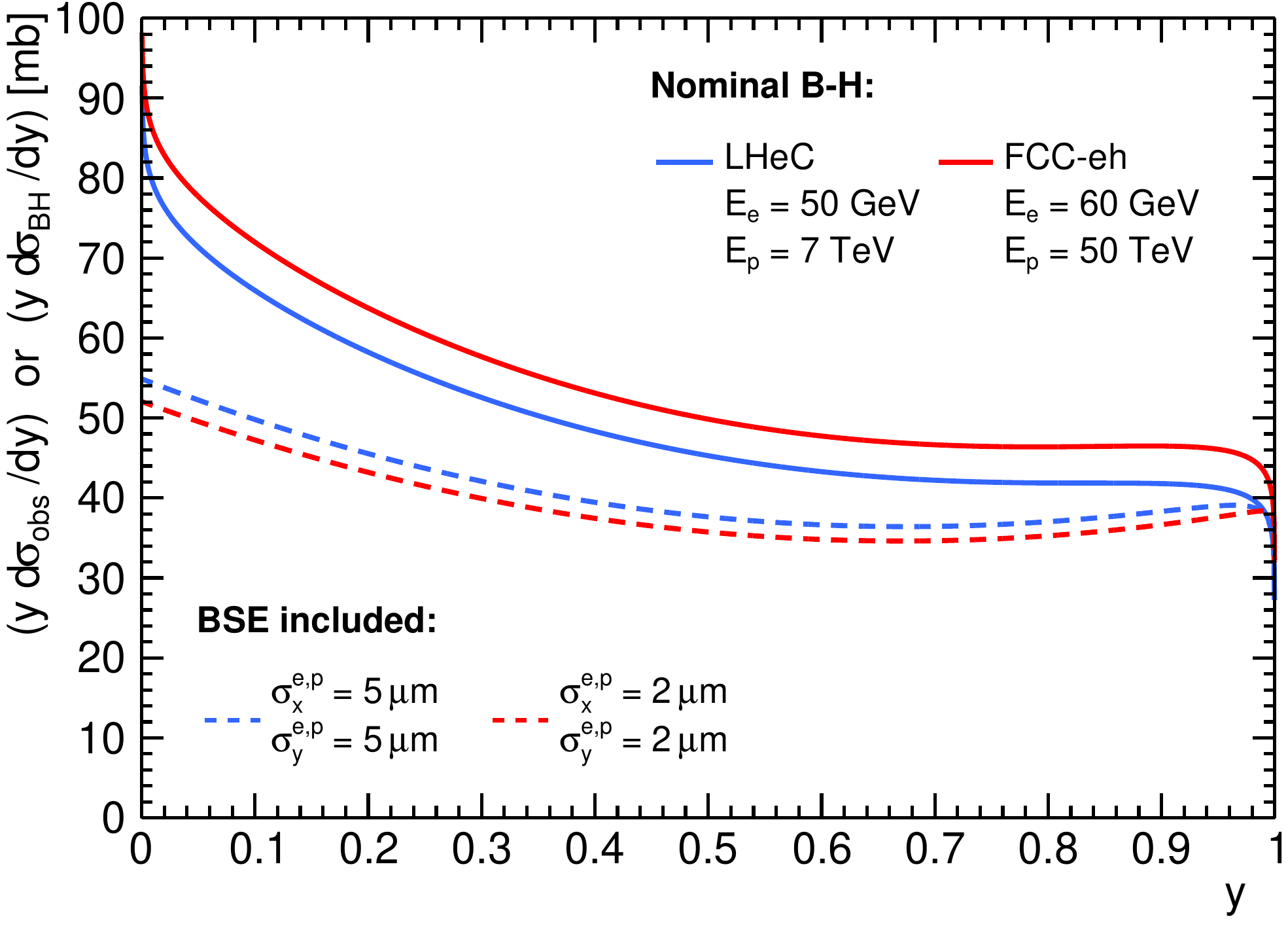}
\caption{Bremsstrahlung spectra $y{\rm d}\sigma/{\rm d}y$ at the LHeC (blue lines) and at the FCC-eh (red lines) -- solid lines are for the Bethe-Heitler nominal case and the dashed when the BSE are included.}
\label{fig:iii}
\end{figure}

At future electron-hadron colliders, as the Large Hadron electron Collider (LHeC) \cite{LHeC:2020van} and the FCC-eh \cite{FCC:2018vvp}, the BSE are large over the whole bremsstrahlung spectra, see Fig.~\ref{fig:iii}. This gives rise to a challenge for precise luminosity measurements, as the BSE corrections should be known with accuracy at a couple of per mille levels.

\begin{figure}[tbph]
\centering
\includegraphics[width=0.75\textwidth]{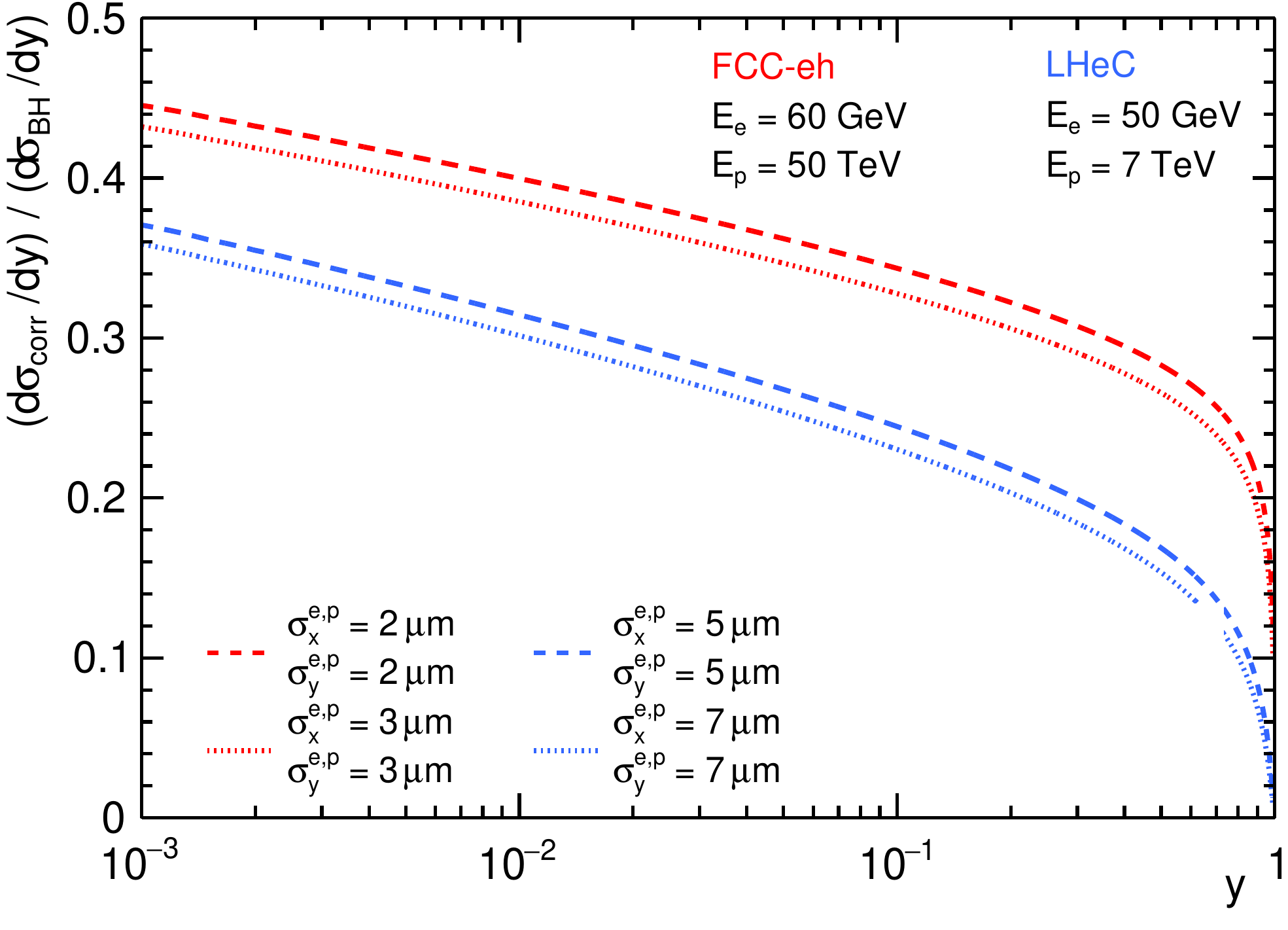}
\caption{Relative size of bremsstrahlung suppression at the LHeC (blue lines) and at the FCC-eh (red lines). Dashed lines are for the nominal beam sizes, and the dotted lines are for the increased beams' lateral sizes as shown in the legend.}
\label{fig:iv}
\end{figure}

\begin{figure}[t!]
\centering
\includegraphics[width=0.75\textwidth]{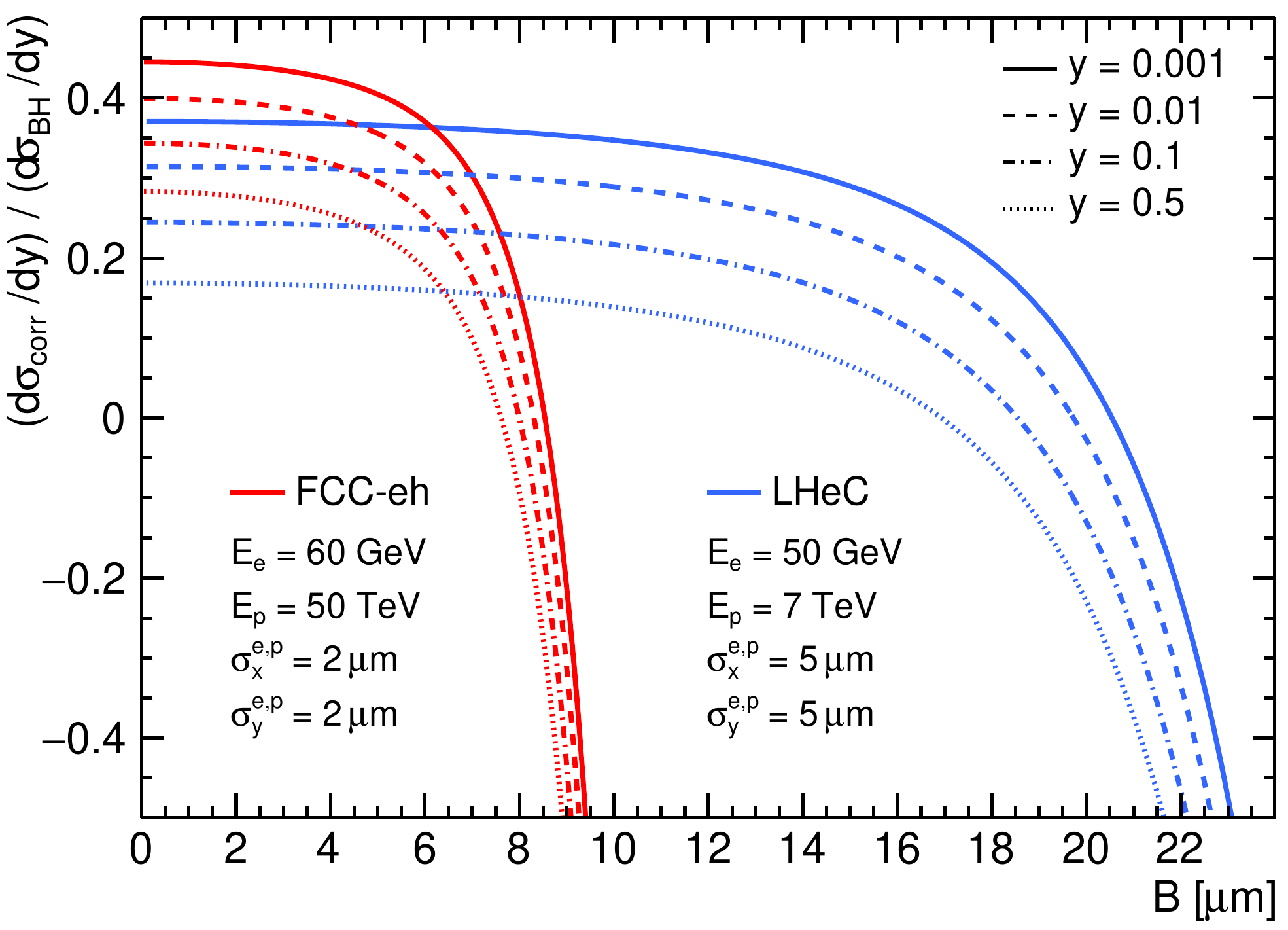}
\caption{Effects of the beam offset $B$ at the FCC-eh (red lines) and LHeC (blue lines) for the nominal beam sizes. Results at different values of $y=E_\gamma/E_e$ are shown with different line styles.}
\label{fig:ii} 
\end{figure}

In particular, a 1\% luminosity precision will require frequent  verifications of Gaussianity of the lateral distribution of bunch particles and continuous measurements of lateral beam sizes with a 5\% precision at least, see Fig. \ref{fig:iv}. The control of beam offsets is less critical, as for small lateral beam displacements, smaller than the lateral beam sizes, the corresponding modifications of BSE are negligible, see Fig. \ref{fig:ii}.

In any case, the experiments at the LHeC and FCC-eh should profit a lot from the experience acquired with the luminosity measurements planned at the EIC, both in the context of proper detector solutions \cite{Piotrzkowski:2021cwj} as well as the precise verification of the BSE predictions~\cite{Piotrzkowski:2020uya}.

As for the FCC-ee case, in Appendix \ref{app_1} a simple formula is provided for quick calculations of the observed cross-sections of bremsstrahlung at the LHeC and FCC-eh.

\subsection{Electron beam losses}
If the LHeC and FCC-eh would be ring-ring colliders then the BSE would be less beneficial for the electron beam lifetimes than at the FCC-ee, as the BSE would be weaker there. However, for both future electron-hadron colliders the linac-ring configurations are being proposed. Hence, as the electron bunches collide only once with the proton ones, the role of bremsstrahlung in the electron beam losses is minimal, see Tab. \ref{tab:iii}.

\begin{table}[tbph]
\centering
{\renewcommand{\arraystretch}{1.2}
\begin{tabular}{|l||c|c|}
\hline
$E_\mathrm{beam}$[GeV] for electrons (protons) & 50 (7000) & 60 (50000)\\
\hline 
$N_b$ [$10^{9}$] per electron bunch & 3 & 3 \\
$L$ [$10^{26}$cm$^{-2}$] per bunch crossing & 3 & 5 \\
$\sigma_{x,y}^{e,p}$ [$\mu$m]  &  5 & 2  \\
$\sigma_{a}$ [mb]  &  219  & 208 \\ 
$\sigma_\mathrm{BH}$ [mb] &  290 & 317 \\ 
\hline
Electron beam fractional loss  & $2.2\times10^{-8}$ & $3.5\times10^{-8}$  \\
\hline
\end{tabular}}
\ \\[10pt]
\caption{The relative electron beam losses due to bremsstrahlung at the LHeC and FCC-eh, for the energy aperture $a=0.01$.}
\label{tab:iii}
\end{table}

\section{Summary}
Bremsstrahlung rates at future high-energy colliders will be strongly affected by small lateral beam sizes, and that will have important consequences. At the FCC-ee, the strong suppression of bremsstrahlung (or radiative Bhabha scattering) will result in a very beneficial phenomenon of doubling the ultimate beam lifetime limits. This is especially relevant for the positron beams at FCC-ee, and potentially also at other colliders under study, as the rate of positron beam refilling, required for high luminosity operations, is one of the major technological problems. On the other hand, at the LHeC and FCC-eh, the bremsstrahlung sensitivity to the beam sizes will pose a big challenge to precise luminosity measurements using this process. Finally, the planned studies of the beam size effects at the EIC will allow to thoroughly verify the theoretical framework used for the BSE calculations.

\bmhead{Acknowledgments}

K. Piotrzkowski is supported by the IDUB programme at the AGH UST and by the NAWA grant BPN/PPO/2021/1/00011.

\begin{appendices}
\section{}
\label{app_1}

At the FCC-ee energies, the observed bremsstrahlung cross-sections approach the asymptotic form, for $y$ not too close to 1. This allows for a simple parameterization:
\begin{equation}
    \frac{\rm{d}\sigma_\mathrm{obs,p}}{{\rm d}y} = \sigma_0\left(1-y+\frac{3}{4}y^2\right)\frac{L}{y}\,C(r),
    \label{eq:2}
\end{equation}
where $\sigma_0=3.093$~mb, $L=\ln{(m_e/\vert Q\vert_\mathrm{cut})}-0.5=13.505,$ with $\vert Q\vert_\mathrm{cut}=0.4227$~eV being the effective "geometrical" cutoff of lateral momentum transfer, and $C(r)=(1+0.074\ln{r})$ is the correction coefficient for a  given ratio $r$ of a vertical beam size to a "reference" beam size of 36~nm, assuming equal bunch sizes for electrons and positrons. For $y\ll 1$ one can simplify to $y\,{\rm d}\sigma_\mathrm{obs,p}/{\rm d}y \simeq \sigma_0 LC(r)= 41.77\times C(r)~[\rm mb]$.

\begin{figure}[tbph]
\centering
\includegraphics[width=0.75\textwidth]{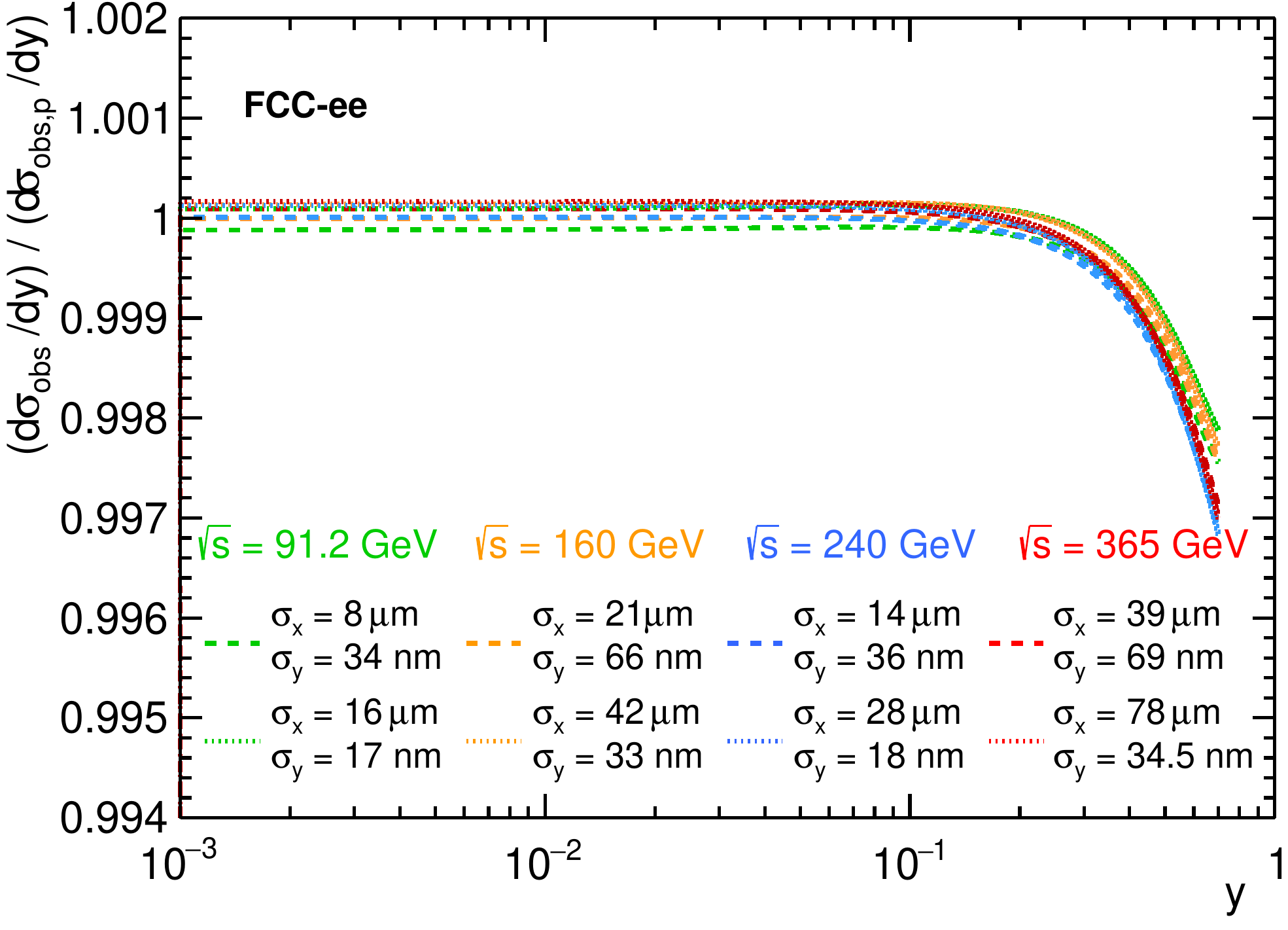}
\caption{Ratios of the FCC-ee cross-sections ${\rm d}\sigma_\mathrm{obs}/{\rm d}y$, as shown in Figs. \ref{fig:1} and \ref{fig:2}, to the  cross-sections ${\rm d}\sigma_\mathrm{obs,p}/{\rm d}y$, according to Eq. \eqref{eq:2}.} 
\label{fig:A1} 
\end{figure}

In Fig. \ref{fig:A1} this parameterization is compared with a number of full calculations, and an excellent agreement is observed, within $\pm0.04\%$ for $y<0.3$. One should note, however, that the precision of Eq. \eqref{eq:2} will deteriorate for the lateral beam sizes largely different from those considered for the FCC-ee at the moment.

The integrated cross-section $\sigma_{a,{\rm p}}=\int^1_{a} {\rm d} \sigma_\mathrm{obs,p}=41.77\times C(r)(a-5/8-3a^2/8-\ln{a})$~[mb], where $a=y_\mathrm{min}$. For $a=0.01$ and $C(r)=1$, one obtains $\sigma_\mathrm{obs,p}=166.67$~mb, to be compared with 166.59 (166.61)~mb from full calculations for $\sigma_y=$~36~nm and $\sigma_x=$~14 (28)~$\upmu$m at $\sqrt{s}=240$ GeV.

\begin{figure}[tbph]
\centering
\includegraphics[width=0.75\textwidth]{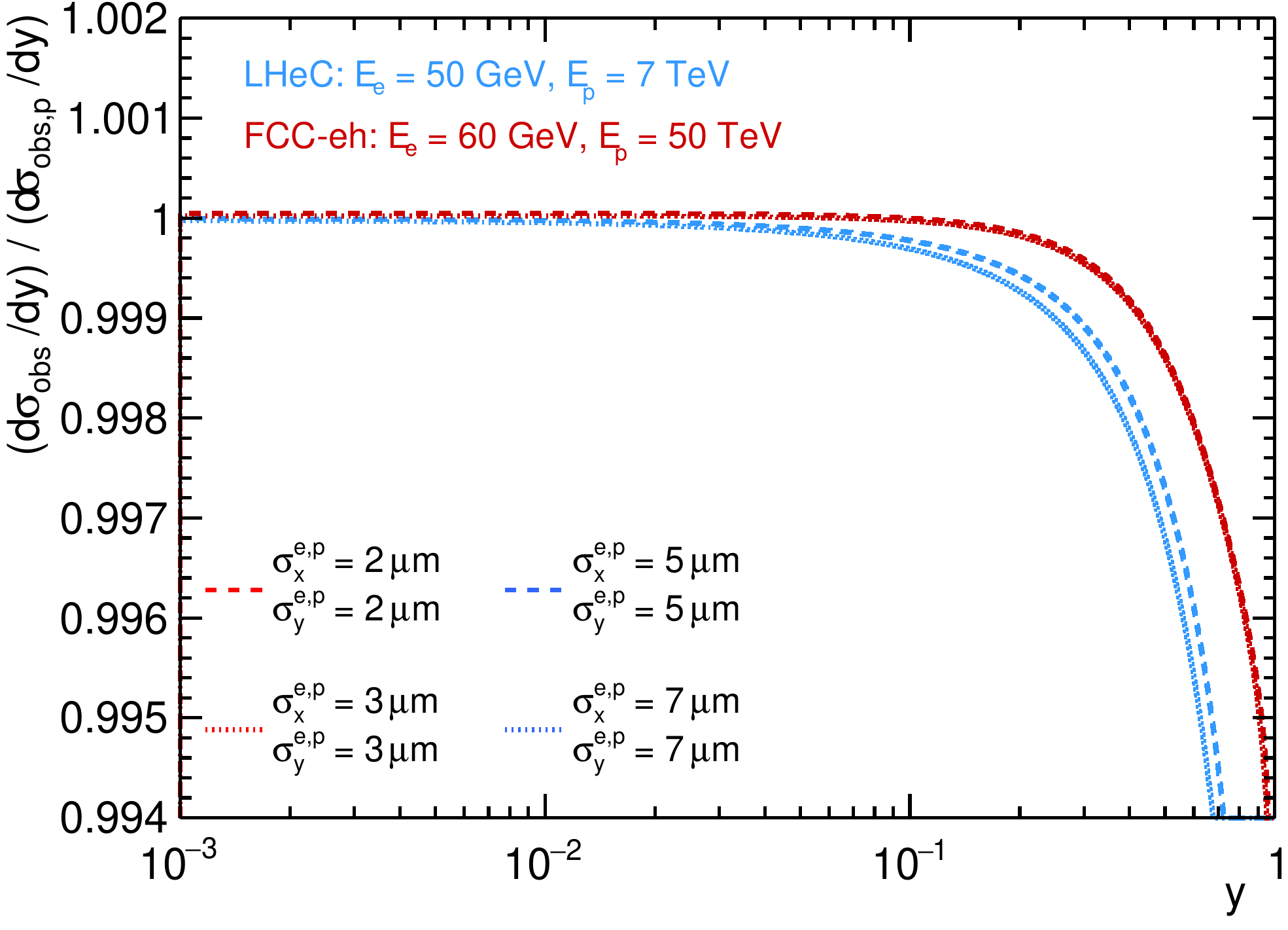}
\caption{Ratios of the FCC-eh and LHeC cross-sections ${\rm d}\sigma_\mathrm{obs}/{\rm d}y$, as shown in Figs. \ref{fig:iii} and \ref{fig:iv}, to the  cross-sections ${\rm d}\sigma_\mathrm{obs,p}/{\rm d}y$, according to Eq. \eqref{eq:2}.} 
\label{fig:A2} 
\end{figure}

For the cases of LHeC and FCC-eh, $\sigma_0 LC(r)= 54.90 \times (1+0.0564\ln{r})~[\rm mb]$, where $r=\sigma/(5~\rm \upmu m$) and $\sigma=\sigma_x=\sigma_y$ for both beams. The agreement between full calculations and the parameterization of BSE cross-sections is very good, as can be seen in Fig. \ref{fig:A2}.

\end{appendices}

\bibliography{bse_lhec_and_fcc}

\end{document}